\documentclass[twocolumn, 
                          showpacs, 
                          nofootinbib, 
                          nobibnotes, 
                          aps,  
                          superscriptaddress, 
                          amssymb, 
                          amsmath, 
                          floatfix, 
                          reprint, 
                          pra, 
                          notitlepage, 
                          showkeys, 
                          10pt]{revtex4-1}

                          \usepackage{amssymb}
\usepackage{amsmath}
\usepackage{graphicx}
\usepackage{dcolumn}
\usepackage[colorlinks=true,breaklinks=true,allcolors=blue]{hyperref}
\usepackage{url}
\usepackage{todonotes}
\usepackage{tikz}
\usepackage{dsfont}
\usepackage{epsfig}
\usepackage{color}
\usepackage{bbold}

\usetikzlibrary{decorations.text}

\usetikzlibrary{shapes.symbols, shapes.arrows, calc, shapes}

\def\equationautorefname~#1\null{Equation~(#1)\null}

\renewcommand{\i}{\mathrm{i}}

\renewcommand{\vec}[1]{\mathbf{#1}}

\newcommand{\eq}[1]{(\ref{eq:#1})}
\newcommand{\Eq}[1]{Eq.\,\eqref{eq:#1}}

\newcommand{\Eqs}[2]{Eqs.~(\ref{eq:#1})--(\ref{eq:#2})}
\newcommand{\Fig}[1]{Fig.~\ref{fig:#1}}

\newcommand{\Sect}[1]{Sect.~\ref{sec:#1}}

\newcommand{\App}[1]{Appendix~\ref{app:#1}}

\newcommand{\Tab}[1]{Table~\ref{tab:#1}}

\newcommand{\1}{\uparrow}

\newcommand{\ket}[1]{|#1\rangle}
\newcommand{\bra}[1]{\langle#1|}
\newcommand{\braket}[2]{\langle#1|#2\rangle}

\makeatletter
\let\cat@comma@active\@empty
\makeatother

\begin{document}

\title{Sampling scheme for neuromorphic simulation of entangled quantum systems}

\author{Stefanie Czischek}
\affiliation{Kirchhoff-Institut f\"ur Physik, Ruprecht-Karls-Universit\"at Heidelberg, Im Neuenheimer Feld 227, 69120 Heidelberg, Germany}
\author{Jan M. Pawlowski}
\affiliation{Institut f\"ur Theoretische Physik, Ruprecht-Karls-Universit\"at Heidelberg, Philosophenweg 16, 69120 Heidelberg, Germany}
\author{Thomas Gasenzer}
\affiliation{Kirchhoff-Institut f\"ur Physik, Ruprecht-Karls-Universit\"at Heidelberg, Im Neuenheimer Feld 227, 69120 Heidelberg, Germany}
\author{Martin G\"arttner}
\affiliation{Kirchhoff-Institut f\"ur Physik, Ruprecht-Karls-Universit\"at Heidelberg, Im Neuenheimer Feld 227, 69120 Heidelberg, Germany}

\date{\today}

\begin{abstract}
  Due to the complexity of the space of quantum many-body states the computation of expectation values by statistical sampling is, in general, a hard task. Neural network representations of such quantum states which can be physically implemented by neuromorphic hardware could enable efficient sampling. A scheme is proposed which leverages this capability to speed up sampling from so-called neural quantum states encoded by a restricted Boltzmann machine. Due to the complex network parameters a direct hardware implementation is not feasible. We overcome this problem by considering a phase reweighting scheme for sampling expectation values of observables. Applying our method to a set of paradigmatic entangled quantum states we find that, in general, the phase-reweighted sampling is subject to a form of sign problem, which renders the sampling computationally costly. The use of neuromorphic chips could allow reducing computation times and thereby extend the range of tractable system sizes.
\end{abstract}

\maketitle

\section{Introduction}
Simulating quantum many-body systems is considered a hard task for classical computers due to the exponentially growing Hilbert space dimension with system size. While some models, like the transverse-field Ising model, can be solved analytically \cite{Pfeuty1970,Calabrese2012,Calabrese2012a,Lieb1961,Sachdev2011}, various approximative simulation methods exist. Among the most successful are the density matrix renormalization group, which is based on tensor network states \cite{Schollwoeck2011,White1992,Vidal2004,Daley2004,Sharma2015,Haegeman2016}, quantum Monte Carlo methods \cite{Sorella2007,Sorella2001}, or semi-classical phase-space methods \cite{Polkovnikov2010,Schachenmayer2015,Schachenmayer2015a,Pucci2016}. 
However, all these methods exploit specific properties of the quantum states.
This limits their range of applicability and requires some a priori knowledge about the system.
Quantum simulation of systems by other quantum systems \cite{Feynman1982a,Bloch2012,Blatt2012,Braun2014,Bernien2017,Zhang2017} as an alternative is plagued by the  fragility of quantum states due to decoherence. 

Here we explore further a new route leading beyond standard von-Neumann architectures which makes use of brain-inspired approaches \cite{Vetra2018,Petrovici2016,Petrovici2016a}. 
Among these, neuromorphic chips emulate neural networks by means of spiking neurons implemented on classical analog hardware. 
These chips have been shown to enable an efficient implementation of sampling from Boltzmann distributions \cite{Petrovici2016,Petrovici2016a,Kungl2019}. 
The sampling on neuromorphic chips can yield a speed-up of at least one order of magnitude compared to classical computers while consuming about three orders of magnitude less energy \cite{Wunderlich2019}. 
The dynamical process performed by the neuromorphic network can be related to Langevin sampling of spin systems \cite{Kades2019} which opens a path to the representation of quantum many-body states on the classical hardware.

Recently, parametrizations of quantum many-body states in terms of a restricted Boltzmann machine (RBM) network topology have been proposed \cite{Carleo2017,Saito2017} and further studied \cite{Gao2017,Deng2017,Nomura2017,Kaubruegger2017,Gao2017,Freitas2018,Carleo2018,Teng2017,Huang2018,Glasser2018,Clark2018,Chen2018,Czischek2018,Pehle2018,Westerhout2019}. 
These RBM representations of quantum states, in general, involve complex network parameters (weights and biases) due to the necessity to account for quantum superposition and interference  \cite{Carleo2017,Torlai2018,Torlai2018a}. 
As a result, the RBM does not resemble a positive definite Boltzmann probability distribution but rather the measure of a Feynman path integral. 
Therefore, standard RBM sampling and training methods, implementable on neuromorphic chips, cannot be applied.

Here we propose a phase reweighting scheme which absorbs the complex phases into the sampled observables \cite{Troyer2005,Anagnostopoulos2002,Nakamura1992,Loh1990,Torlai2019b,Hangleiter2019}, while the remaining amplitudes give a Boltzmann distribution of the spin configurations. 
This enables an implementation of a sampling process from this distribution on neuromorphic hardware to approximately evaluate expectation values according to a quantum Monte Carlo method. 

Moreover, an extension of the parametrization to deep Boltzmann machines with multiple hidden layers becomes possible.
In contrast to two-layer RBMs with complex weights which require the analytic summation of the hidden spins \cite{Carleo2017}, these networks can be Gibbs sampled.
This is crucial since an RBM with a single hidden layer allows to extract directly only diagonal spin observables from the sampled visible spin configurations. 
Here we devise an extension to deep networks which makes the calculation of arbitrary expectation values of spin operators possible from the visible layer of the network. 

In general, the advantages of the reweighting scheme come at the expense of a form of sign problem which renders the sampling costly. 
However, we argue that neuromorphic hardware implementations could still provide a proportional speedup compared to standard architectures \cite{Wunderlich2019}.
To benchmark our method, we consider the transverse-field Ising model (TFIM), which is a one-dimensional spin-1/2 chain with nearest-neighbor Ising interactions in a transverse field. The model describes a quantum phase transition between a para- and a ferromagnetic phase, controlled by the relative strength of the interactions and the transverse-field strength. We apply the deep Boltzmann machine representation to perform measurements of the order parameter and correlations in the ground state of the TFIM at the quantum phase transition, which is known to exhibit non-trivial quantum correlations.

To further analyze the ability of the RBM representation to capture quantum mechanical effects, we consider spin systems in strongly entangled states. As paradigmatic examples we choose the Bell state of two spins \cite{Bell1964,Bell1966,Bell2004}, as well as its generalization to larger spin systems, the Greenberger-Horne-Zeilinger (GHZ) states \cite{Greenberger1989}.

If the network parameters are purely real, which is the case for measurements in the computational basis ($z$-basis) for the ground state of the TFIM, we find sampling to be efficient, in the sense that the number of random samples needed in order for the expectation values to be converged is almost independent of the system size. Measurements in other bases require the extension to a deep network. For Bell and GHZ states, where complex network parameters are involved, the number of samples required to reach converged results scales exponentially with the size of the network. This is a manifestation of the sign problem \cite{Nakamura1992, Troyer2005, Anagnostopoulos2002, Loh1990, Broecker2017,Torlai2019,Hangleiter2019}, which is due to the fact that the phases that need to be averaged in the phase reweighting scheme fluctuate heavily. This leads to strongly growing variances in the statistical sampling process (see \Sect{results-1}). As a result, our method can be used to encode quantum states of limited size, also maximally entangled ones, in classical networks and sample from them, which can be realized  efficiently by neuromorphic hardware. 

Applications of RBM representation of quantum states include the simulation of ground states and dynamics in closed and open quantum many-body systems \cite{Carleo2017,Saito2017,Hartmann2019,Nagy2019,Vincentini2019,Yoshioka2019,Lu2019,Gao2017,Deng2017,Nomura2017,Hartmann2019,Nagy2019,Vincentini2019,Yoshioka2019,Kaubruegger2017,Freitas2018,Carleo2018,Teng2017,Czischek2018,Westerhout2019,Gardas2018} and efficient state tomography by learning the parameters of the state from experimental data \cite{Carrasquilla2019,Torlai2018,Torlai2019}.
All these applications could benefit from the use of neuromorphic hardware to more efficiently perform learning and sampling tasks. 

After having introduced the RBM ansatz for parametrizing quantum states of many-spin systems (Sec.~\ref{sec:RBMstate}), we present the phase reweighting scheme (Sec.~\ref{sec:reweight}) and extensions to deep networks for measuring off-diagonal observables (Sec.~\ref{sec:DNN}). 
In Secs. \ref{sec:TFIM} and \ref{sec:Bell}, we apply the method to ground states of the TFIM and Bell and GHZ states, respectively, analyzing convergence and sampling efficiency. We draw conclusions in Sec.~\ref{sec:conclusion}.

\section{Representing Quantum States with Restricted Boltzmann Machines}\label{sec:RBMstate}
The state vector of an array of $N$ spin-1/2 objects can be expressed in terms of product basis states $\ket{\vec{v}^z}=\ket{v_1^z}\otimes\dots \otimes \ket{v_N^z}$, where $v_i^z=\pm 1$ and $\ket{v_i^z}$ are the eigenstates of the Pauli operator $\sigma^z$. The many-body wave-function is thus represented by complex coefficients $c_{\vec{v}^z}\in\mathbb{C}$ as
\begin{align}
  \label{eq:1}
  \ket{\Psi}=&\sum_{\left\{\vec{v}^z\right\}}c_{\vec{v}^z}\ket{\vec{v}^z},
\end{align}
where the sum runs over all basis states. The number of coefficients needed is $2^N$, thus scaling exponentially in system size.
In order to reduce the number of parameters needed to represent the quantum state to a tractable size, the coefficients $c_{\vec{v}^z}$ can be parametrized by means of an artificial neural network, specifically of a restricted Boltzmann machine (RBM) \cite{Torlai2016, Carleo2017, Saito2017}.

\begin{figure}
  \centering
  \includegraphics{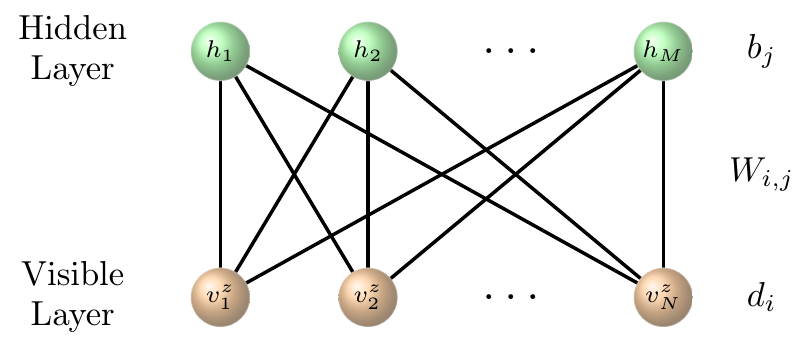}
  \caption{Restricted Boltzmann machine (RBM) network with $N$ visible neurons $\vec{v}^z=(v_1^z,\dots,v_N^z)$ and $M$ hidden neurons $\vec{h}$, cf.~Eqs.~\eq{2}, \eq{3}. 
  Each visible neuron $v_i^z$ is connected to each hidden neuron $h_j$ via the weight $W_{i,j}$ and has a bias $d_i$, while the hidden neurons $h_j$ have biases $b_j$.}
  \label{fig:1}
\end{figure}

An RBM consists of $N$ visible and $M$ hidden binary neurons, $v_i^z,h_j\in\{\pm1\}$, where the visible neurons correspond to the physical spins in the quantum state representation. Each visible neuron $v_i^z$ is connected to each hidden neuron $h_j$ via a weight $W_{i,j}$ and each neuron has an additional bias, $d_i$ for the visible and $b_j$ for the hidden neurons, as illustrated in \Fig{1}. This yields the network energy function
\begin{align}
  \label{eq:2}
  \begin{split}
    E_{\mathrm{RBM}}\left(\vec{v}^z,\vec{h};\mathcal{W}\right)=&\ -\sum_{i=1}^N\sum_{j=1}^Mv_i^zW_{i,j}h_j-\sum_{i=1}^Nv_i^zd_i\\
    &\ -\sum_{j=1}^Mh_jb_j,
  \end{split}
\end{align}
with the set of all weights and biases $\mathcal{W}=(\vec{d},\vec{b},W)$ \cite{Hinton2012}.\\
An RBM-based parametrization of the unnormalized coefficients $c_{\vec{v}^z}$ can be defined as \cite{Torlai2016, Carleo2017}
\begin{align}
  \label{eq:3}
  c_{\vec{v}^z}\left(\mathcal{W}\right)=&\ \sum_{\left\{\vec{h}\right\}}\mathrm{exp}\left[-E_{\mathrm{RBM}}\left(\vec{v}^z,\vec{h};\mathcal{W}\right)\right].                                         
\end{align}
While for real-valued network parameters this corresponds to the marginal of a Boltzmann distribution over all neurons of the network, the weights and biases in general need to be complex in order to account for the complex basis projections $c_{\vec{v}^z}$ of a quantum state \cite{Torlai2016, Carleo2017}.

An unnormalized probability distribution over the visible neurons is instead given by $|c_{\vec{v}^z}\left(\mathcal{W}\right)|^2$, which can be evaluated after the summation over the binary hidden neurons is carried out, giving \cite{Carleo2017}
\begin{align}
  \label{eq:4}
  c_{\vec{v}^z}\left(\mathcal{W}\right)=&\ \mathrm{exp}\left[\sum_{i=1}^Nd_iv_i^z\right]\prod_{j=1}^M2\mathrm{cosh}\left[\sum_{i=1}^Nv_i^zW_{i,j}+b_j\right] .
\end{align}
With this, expectation values of operators that are diagonal in the chosen basis [$z$-basis, cf.\ \Eq{1}], $\mathcal{O}^{\mathrm{diag}}$, can be written as
\begin{align}
  \label{eq:5}
  \begin{split}
    \left<\mathcal{O}^{\mathrm{diag}}\right>=&\ \frac{1}{Z\left(\mathcal{W}\right)}\sum_{\left\{\vec{v}^z\right\}}\mathcal{O}^{\mathrm{diag}}\left(\vec{v}^z\right)\left|c_{\vec{v}^z}\left(\mathcal{W}\right)\right|^2\\
    \approx&\ \frac{1}{Q}\sum_{q=1}^Q\mathcal{O}^{\mathrm{diag}}\left(\vec{v}_q^z\right),
  \end{split}
\end{align}
normalized by
\begin{align}
  \label{eq:7}
  Z\left(\mathcal{W}\right)=&\sum_{\left\{\vec{v}^z\right\}}\left|c_{\vec{v}^z}\left(\mathcal{W}\right)\right|^2.
\end{align}
In the second line of \Eq{5} we approximate the mean value in terms of a sum over $Q$ samples of visible spin configurations $\vec{v}^z$ drawn according to the probability distribution $|c_{\vec{v}^z}(\mathcal{W})|^2$ via a Metropolis-Hastings algorithm and evaluate $\mathcal{O}^{\mathrm{diag}}(\vec{v}_q^z)$ on these samples \cite{Carleo2017}. The sampling error can be estimated via the variance, as further discussed in \Sect{results}.

Non-diagonal operators, such as the total magnetization $\sum_i \sigma_i^x$, can  be evaluated similarly, exploiting their sparsity \cite{Carleo2017}:
\begin{align}
  \label{eq:23}
  \begin{split}
    \left<\mathcal{O}\right>=&\ \frac{1}{Z\left(\mathcal{W}\right)}\sum_{\left\{\vec{v}^z\right\}}\sum_{\left\{\tilde{\vec{v}}^z\right\}}\left<\tilde{\vec{v}}^z\left|\mathcal{O}\right|\vec{v}^z\right>c_{\vec{v}^z}\left(\mathcal{W}\right)c_{\tilde{\vec{v}}^z}^{*}\left(\mathcal{W}\right)\\
    =&\ \frac{1}{Z\left(\mathcal{W}\right)}\sum_{\left\{\vec{v}^z\right\}}\mathcal{O}^{\mathrm{loc}}\left(\vec{v}^z\right)\left|c_{\vec{v}^z}\left(\mathcal{W}\right)\right|^2\\
    \approx&\ \frac{1}{Q}\sum_{q=1}^Q\mathcal{O}^{\mathrm{loc}}\left(\vec{v}^z_q\right),
  \end{split}
\end{align}
where we introduce the local operator
\begin{align}
  \label{eq:35}
  \mathcal{O}^{\mathrm{loc}}\left(\vec{v}^z\right)=&\sum_{\left\{\tilde{\vec{v}}^z\right\}}\left<\tilde{\vec{v}}^z\left|\mathcal{O}\right|\vec{v}^z\right>\frac{c_{\tilde{\vec{v}}^z}^{*}\left(\mathcal{W}\right)}{c_{\vec{v}^z}^{*}\left(\mathcal{W}\right)},
\end{align}
with the star denoting complex conjugation. This expression can be evaluated efficiently if $\mathcal{O}$ is sparse, i.\,e. $\bra{\vec{v}^z}\mathcal{O}\ket{\vec{\tilde v}^z}$ is only non-vanishing for a number of matrix elements that scales polynomially in the number of spins. Physically relevant observables have this property.

The weights and biases in the RBM parametrization are variational parameters which can be adapted to represent a desired wave function. These weights can be found via a variational ansatz, where commonly a stochastic reconfiguration method is used to find ground state representations via energy minimization \cite{Carleo2017}. In some cases the network parameters can be found analytically as is the case for Bell and GHZ states which we will use below.

In the case of complex network parameters, Eq.~\eqref{eq:3} can no longer be interpreted as the marginal of a Boltzmann distribution. The exponential factors that are summed are complex and thus do not represent probabilities. In particular, this means that the intuitive procedure for evaluating observables -- sampling configurations of all neurons in the network, including hidden ones, and then averaging the values of the observables obtained from the states of the visible neurons -- is not applicable.

This motivates us to reformulate the procedure of evaluating observables in a way that allows sampling from a Boltzmann distribution while complex phases are absorbed into the diagonal elements of the considered observable. On the one hand, this enables an extension to deep neural networks with multiple hidden layers. On the other hand, an implementation on neuromorphic hardware setups becomes possible, which is known to efficiently sample from Boltzmann distributions \cite{Petrovici2016a,Petrovici2016,Kungl2019}.

\section{Phase Reweighting Scheme}\label{sec:reweight}
To enable a sampling of visible and hidden neurons in the complex RBM from Boltzmann distributions, we consider the exponential of the network energy stated in \Eq{2}. We split the weights and biases into real and imaginary parts,
\begin{align}
  \label{eq:8}
  \begin{split}
    \mathrm{exp}&\left[-E_{\mathrm{RBM}}\left(\vec{v}^z,\vec{h};\mathcal{W}\right)\right]\\
    =&\ \mathrm{exp}\left[\sum_{i=1}^N\sum_{j=1}^Mv_i^zW_{i,j}^{\mathrm{R}}h_j+\sum_{i=1}^Nd_i^{\mathrm{R}}v_i^z+\sum_{j=1}^Mb_j^{\mathrm{R}}h_j\right]\\
    &\times\mathrm{exp}\left[i\left(\sum_{i=1}^N\sum_{j=1}^Mv_i^zW_{i,j}^{\mathrm{I}}h_j+\sum_{i=1}^Nd_i^{\mathrm{I}}v_i^z+\sum_{j=1}^Mb_j^{\mathrm{I}}h_j\right)\right]\\
    =:&\ \tilde{P}\left(\vec{v}^z,\vec{h};\mathcal{W}^{\mathrm{R}}\right)e^{i\tilde{\varphi}\left(\vec{v}^z,\vec{h};\mathcal{W}^{\mathrm{I}}\right)},
  \end{split}
\end{align}
with $\mathcal{W}=\mathcal{W}^{\mathrm{R}}+i\mathcal{W}^{\mathrm{I}}$.
Here we introduce the probability distribution $\tilde{P}(\vec{v}^z,\vec{h};\mathcal{W}^{\mathrm{R}})$ yielding a Boltzmann distribution over visible and hidden neurons, and the phase $\tilde{\varphi}(\vec{v}^z,\vec{h};\mathcal{W}^{\mathrm{I}})$.

The basis expansion coefficients can then be expressed as
\begin{align}
  \label{eq:9}
  c_{\vec{v}^z}\left(\mathcal{W}\right)=&\sum_{\left\{\vec{h}\right\}}\tilde{P}\left(\vec{v}^z,\vec{h};\mathcal{W}^{\mathrm{R}}\right)e^{i\tilde{\varphi}\left(\vec{v}^z,\vec{h};\mathcal{W}^{\mathrm{I}}\right)},
\end{align}
according to \Eq{3}. Substituting \Eq{9} into \Eq{5}, expectation values of diagonal operators in the $z$-basis, $\mathcal{O}^{\mathrm{diag}}$, can be expressed as
\begin{align}
  \label{eq:10}
  \begin{split}
    \left<\mathcal{O}^{\mathrm{diag}}\right>=&\ \frac{1}{Z\left(\mathcal{W}\right)}\sum_{\left\{\vphantom{\tilde{\vec{h}}}\vec{v}^z\right\}}\sum_{\left\{\vec{h},\tilde{\vec{h}}\right\}}\left[\mathcal{O}^{\mathrm{diag}}\left(\vec{v}^z\right)e^{i\varphi\left(\vec{v}^z,\vec{h},\tilde{\vec{h}};\mathcal{W}^{\mathrm{I}}\right)}\right]\\
    &\hphantom{\frac{1}{Z\left(\mathcal{W}\right)}\sum_{\left\{\vec{v}^z\right\}}\sum_{\left\{\vec{h}\right\}}\sum_{\left\{\tilde{\vec{h}}\right\}}}\times P\left(\vec{v}^z,\vec{h},\tilde{\vec{h}};\mathcal{W}^{\mathrm{R}}\right),
  \end{split}\\
  \label{eq:11}
  \begin{split}
    Z\left(\mathcal{W}\right)=&\sum_{\left\{\vec{v}^z\vphantom{\tilde{\vec{h}}}\right\}}\sum_{\left\{\vec{h},\tilde{\vec{h}}\right\}}e^{i\varphi\left(\vec{v}^z,\vec{h},\tilde{\vec{h}};\mathcal{W}^{\mathrm{I}}\right)}P\left(\vec{v}^z,\vec{h},\tilde{\vec{h}};\mathcal{W}^{\mathrm{R}}\right),
  \end{split}
\end{align}
where we introduce
\begin{align}
  \label{eq:12}
  \varphi\left(\vec{v}^z,\vec{h},\tilde{\vec{h}};\mathcal{W}^{\mathrm{I}}\right)=:&\ \tilde{\varphi}\left(\vec{v}^z,\vec{h};\mathcal{W}^{\mathrm{I}}\right) - \tilde{\varphi}\left(\vec{v}^z,\tilde{\vec{h}};\mathcal{W}^{\mathrm{I}}\right),\\
  \label{eq:13}
  P\left(\vec{v}^z,\vec{h},\tilde{\vec{h}};\mathcal{W}^{\mathrm{R}}\right)=:&\ \tilde{P}\left(\vec{v}^z,\vec{h};\mathcal{W}^{\mathrm{R}}\right)\tilde{P}\left(\vec{v}^z,\tilde{\vec{h}};\mathcal{W}^{\mathrm{R}}\right).
\end{align}
It turns out that the hidden neuron configurations are summed over twice (sums over $\vec{h}$ and $\tilde{\vec{h}}$) when calculating expectation values. These sums originate from the coefficients of bra- and ket-states, respectively \cite{Hartmann2019}.

As $P(\vec{v}^z,\vec{h},\tilde{\vec{h}};\mathcal{W}^{\mathrm{R}})$ yields a Boltzmann distribution over the visible and hidden neurons, the sums in \Eqs{10}{11} can be approximated by summing over samples drawn from these probability distributions. We use standard block Gibbs sampling \cite{Hinton2012}.
In contrast to \Eq{5}, not only the operator itself needs to be evaluated for each sample, but also the phase $\varphi(\vec{v}^z,\vec{h},\tilde{\vec{h}};\mathcal{W}^{\mathrm{I}})$. Furthermore, this phase appears in the normalization factor $Z(\mathcal{W})$ [cf.\ Eq.~\eqref{eq:11}].

The observable is thus reweighted with a complex phase in the evaluation of expectation values, which is why one refers to this ansatz as a phase reweighting scheme. This is a commonly used method in quantum Monte Carlo approaches \cite{Troyer2005,Anagnostopoulos2002,Nakamura1992,Loh1990,Torlai2019,Hangleiter2019}.
Note that this method often suffers from a sign problem. If the phases fluctuate heavily they can cancel each other, resulting in an uncontrolled growth of the variance of the sampled quantity, which in turn requires an exponentially growing number of samples \cite{Nakamura1992, Troyer2005, Anagnostopoulos2002, Loh1990}. A more quantitative account of this will be given in \Sect{results}.

\section{Measuring in Different Bases}\label{sec:DNN}
Having introduced the phase reweighting scheme to evaluate expectation values of diagonal operators by sampling from a Boltzmann distribution, we now derive a scheme for measuring non-diagonal operators that eliminates the use of local operators. This overcomes the problem that efficient evaluation of local operators requires them to be sparse and that their matrix elements need to be evaluated explicitly. More importantly, only the states of the visible spins and the phases associated with a given sample state need to be evaluated for the samples drawn from a Boltzmann distribution. This enables the use of neuromorphic architectures for performing the sampling.

\begin{figure}
  \centering
  \includegraphics[width=\linewidth]{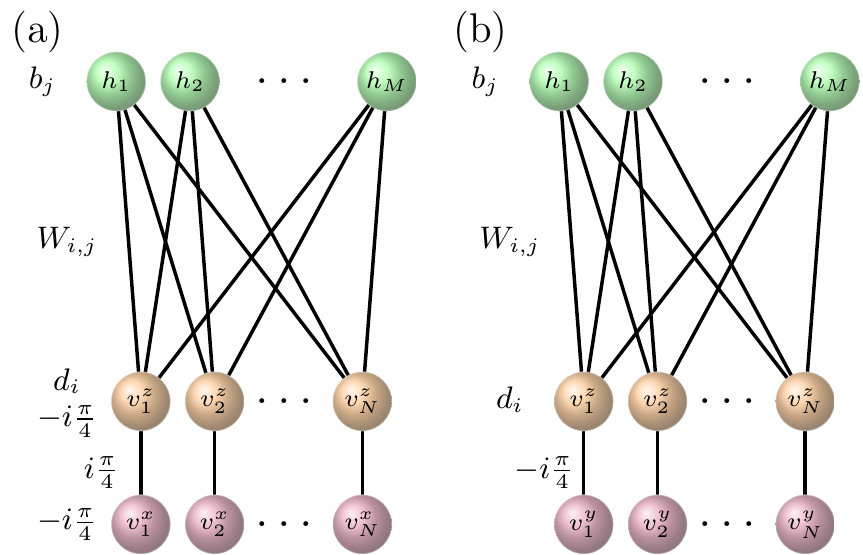}
  \caption{Setup of the deep neural network parametrization to perform measurements in different spin bases. 
  Panel (a) shows the extension of the RBM with visible neurons $v_i^z$, hidden neurons $h_j$, biases $d_i$, $b_j$ and connecting weights $W_{i,j}$, cf.~\Fig{1}, to a deep neural network (dNN) to represent the state in the $x$-basis with visible neurons $v_i^x$, where the neurons $v_i^z$ turn into hidden neurons which are summed over. 
  $v_i^x$ and $v_i^z$ are connected by weights $i\pi/4$, and biases for $v_i^x$ as well as imaginary parts of the biases for $v_i^z$ appear. 
  Panel (b) shows the corresponding dNN to perform measurements in the $y$-basis, where the visible neurons $v_i^y$ are added instead of $v_i^x$ in panel (a). 
  Here only connecting weights between $v_i^y$ and $v_i^z$ are required, but no further biases appear.}
  \label{fig:3}
\end{figure}

Any hermitian operator can be decomposed into a Pauli string as
\begin{equation}
 \mathcal{O} = \sum_{\left\{\boldsymbol\alpha\right\}} D_{\boldsymbol\alpha} \sigma_1^{\alpha_1}\otimes \ldots\otimes \sigma_N^{\alpha_N}
\end{equation}
where $\alpha_i\in \{0,1,2,3\}$ and $\sigma_i^{\alpha_i}$ are Pauli operators acting on spin $i$ with $\sigma_i^0=\mathbb{1}$ and $D_{\boldsymbol\alpha}$ are expansion coefficients. Here, we will only consider product operators or Pauli strings, such as the magnetization of a single spin $\sigma_i^{\alpha_i}$ (not writing out identities) or correlations between two spins $\sigma_i^{\alpha_{i\vphantom{j}}}\sigma_j^{\alpha_j}$, but generalization to arbitrary operators is straight forward. 

An operator is diagonal in the $z$-basis if it only involves $\sigma^3=\sigma^z$-operators (and identities).
A non-diagonal spin operator, which involves $\sigma^x$ and/or $\sigma^y$, can be evaluated by rotating the spins locally into the basis in which the operators acting on the corresponding spins are diagonal. In the following we will show that this procedure can be encoded as an additional layer in the neural network representation of the state. 

We first consider the Pauli-operator $\sigma_i^x$ acting on a single spin $i$. The eigenstates of this operator can be obtained from the $z$-basis states by applying a rotation,
\begin{align}
  \label{eq:14}
  \ket{v_i^x}=&\sum_{\left\{v_i^z\right\}}u_{z\rightarrow x}\left(v_i^x,v_i^z\right)\ket{v_i^z},
\end{align}
with the entries $u_{z\rightarrow x}(v_i^x,v_i^z)=\braket{v_i^z}{v_i^x}$ of the unnormalized rotation matrix
\begin{align}
  \label{eq:15}
  U_{z\rightarrow x}=&\left[\begin{matrix}1&1\\1&-1\end{matrix}\right].
\end{align}
These entries can be written as an exponential function,
\begin{align}
  \label{eq:16}
  u_{z\rightarrow x}\left(v_i^x,v_i^z\right)=&\ \mathrm{exp}\left[i\frac{\pi}{4}\left(v_i^xv_i^z-v_i^x-v_i^z+1\right)\right],
\end{align}
taking the form of the exponential of an RBM network energy, similar to \Eq{2}.
In order to evaluate the expectation value of $\sigma_i^x$ we can equivalently apply a rotation to spin $i$ and then evaluate the diagonal observable $\sigma_i^z$. This means that we replace the single-particle basis states $\ket{v_i^z}$ appearing in \Eq{1} by \Eq{14}, resulting in an additional summation over $v_i^z$. This additional summation means in our network representation that $v_i^z$ has become a hidden neuron. Since the involved matrix elements can be written as exponential factors of the required form \Eq{16}, they just contribute to the overall network energy as any other weights in the network. We hence end up with a deep neural network (dNN) with two hidden layers. If we are interested in measuring the $x$-magnetization of all spins or correlations $\sigma_i^x\sigma_j^x$, we would add an additional connection to every visible spin as illustrated in Fig.~\ref{fig:3}(a).

To summarize, we can parametrize the state of a spin in the $x$-basis by introducing a neuron $v_i^x$ in the RBM parametrization, which is connected to $v_i^z$ via a weight $i\pi/4$ and has a bias of $-i\pi/4$, while $v_i^z$ gets an additional bias of $-i\pi/4$. The overall bias of $i\pi/4$ appearing in \Eq{16} can be neglected as it is an irrelevant global phase factor.

An analogous expression can be derived for a local rotation of the spin state into the $y$-basis, so that also a neuron $v_i^y$ can be connected to $v_i^z$, enabling a measurement of the Pauli-operator $\sigma_i^y$, as illustrated in \Fig{3}(b). The $v_i^z$ neuron again turns into a hidden neuron, yielding the transformation
\begin{align}
  \label{eq:32}
  \ket{v_i^y}=&\sum_{\left\{v_i^z\right\}}u_{z\rightarrow y}\left(v_i^y,v_i^z\right)\ket{v_i^z},
\end{align}
with elements
\begin{align}
  \label{eq:33}
  u_{z\rightarrow y}\left(v_i^y,v_i^z\right)=\ &\mathrm{exp}\left[i\frac{\pi}{4}\left(1-v_i^yv_i^z\right)\right],
\end{align}
of the rotation matrix
\begin{align}
  \label{eq:34}
  U_{z\rightarrow y}=&\left[\begin{matrix}1&i\\i&1\end{matrix}\right].
\end{align}
With this ansatz, any desired Pauli string operator can be measured using the phase reweighting scheme on the corresponding dNN. The network representation can always be set up to represent the local spins in the basis where the applied operator becomes diagonal. Notice that the added network parameters are purely imaginary such that the resulting dNN will always contain complex parameters. We saw in \Eqs{12}{13} that the evaluation of observables requires to sum twice over all hidden neurons. This will now be the case for all spins $v_i^z$ that get connected to an additional spin $v_i^x$ or $v_i^y$.

\section{Ground States of the Transverse-Field Ising Model}\label{sec:TFIM}
Having introduced the dNN setup with the phase reweighting scheme, we benchmark the approach on the ground state of the transverse-field Ising model (TFIM) at the quantum critical point.
\subsection{Model and dNN Representation}
The one-dimensional TFIM is an integrable model defined on a spin-1/2 chain with $N$ sites via the Hamiltonian
\begin{align}
  \label{eq:6}
  H_{\mathrm{TFIM}}=-J\sum_{i=1}^N\sigma_i^z\sigma_{\left(i+1\right)\mathrm{mod} N}^z-h\sum_{i=1}^N\sigma_i^x,
\end{align}
with $(i+1)\mathrm{mod}N$ denoting a modulo-$N$ calculation, i.e., we choose periodic boundary conditions. The system undergoes a quantum phase transition at the quantum critical point reached for the transverse magnetic field strength $h=h_{\mathrm{c}}=J$. In the following, we fix the energy scale by setting $J=1$.
The model is integrable and can be solved in terms of a Jordan-Wigner-fermionization \cite{Pfeuty1970, Calabrese2012, Calabrese2012a, Lieb1961, Sachdev2011, Karl2017}. It has been studied in great detail and is a common choice to benchmark approximative analytical and numerical methods.

To represent the ground state in the RBM parametrization, the corresponding weights can be found variationally by using stochastic reconfiguration minimizing the system energy \cite{Carleo2017}. In the following we apply this variational ansatz to find weights and biases representing the ground state at the quantum critical point. Subsequently, given the state representation, we fix the weights and biases and apply the phase reweighting scheme to benchmark its performance when applied in the dNN approach.

\begin{figure*}
  \centering
  \includegraphics[width=\textwidth]{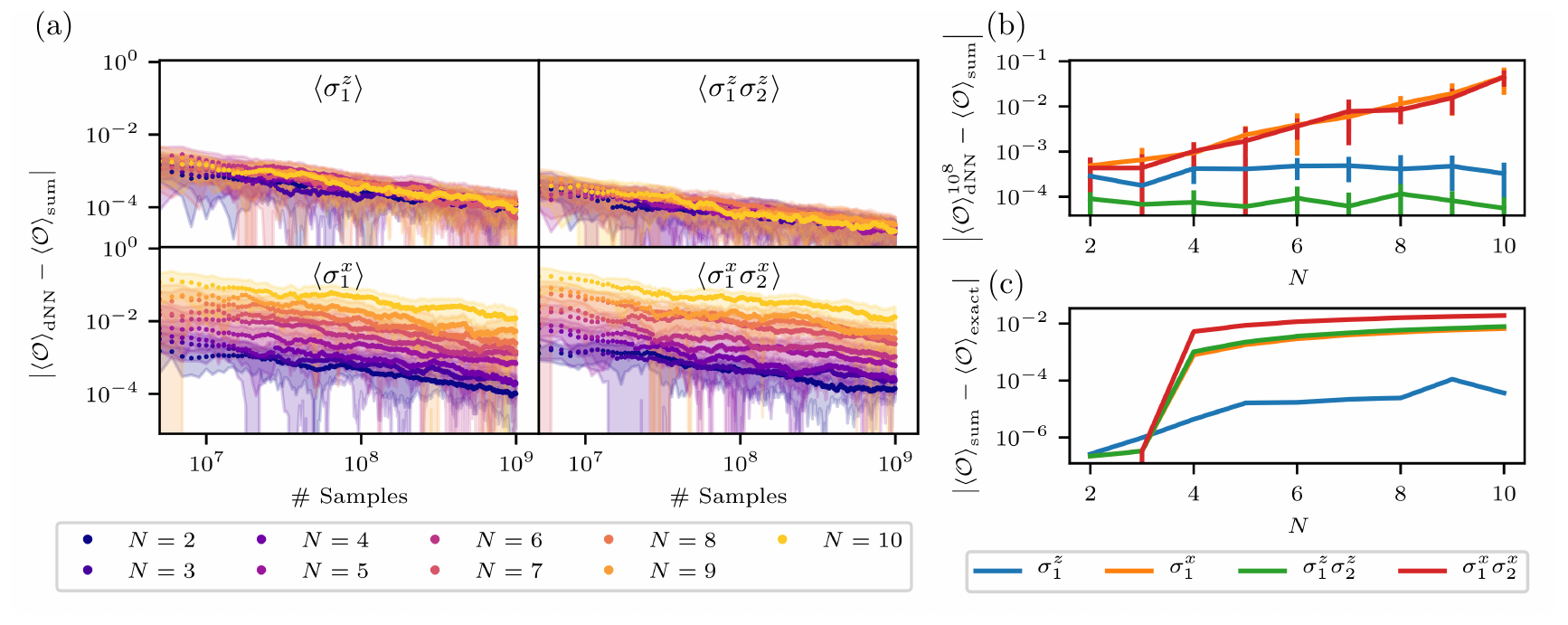}
  \caption{RBM representation of the TFIM ground state at the quantum critical point:   
  Sampling and representation errors for the phase reweighting scheme. 
  Magnetizations $\langle\sigma_{1}^{k}\rangle$ of the first spin and correlations $\langle\sigma_{1}^{k}\sigma_{2}^{k}\rangle$ between the first two spins in a TFIM chain of $N$ spins, in $k$-direction, $k=z,x$, have been evaluated.
  Panel (a) shows the absolute deviation of the observables evaluated with the phase reweighting method, $\langle\mathcal{O}\rangle_{\mathrm{dNN}}$, from the respective value obtained by explicitly summing over all states, $\langle\mathcal{O}\rangle_{\mathrm{sum}}$, as a function of the sample size. 
  The number of spins in the chain is varied from $N=2$ to $N=10$, indicated by the colors. 
  The sampling is run ten times for each system size and the results are averaged, where shaded regions denote statistical fluctuations. 
  While we do not find a dependence on the system size for measurements in the $z$-basis (upper row), we find increasing deviations with larger $N$ for measurements in the $x$-basis (lower row). 
  In both cases the error decays only slowly and a huge amount of samples is considered.  
Panel (b) shows the same data evaluated at a fixed sample size of $10^8$, as a function of system size $N$, as indicated in the legend at the bottom. 
Error bars denote statistical fluctuations within averaging over ten runs. 
We find the accuracy of the measurements in the $z$-basis to be approximately constant, while the deviations grow exponentially for measurements in the $x$-basis. 
The latter indicate the existence of a sign problem which limits the method to small system sizes.
Panel (c) shows the absolute deviation between the observables as calculated in the RBM parametrization but by summing over all configurations, $\langle\mathcal{O}\rangle_{\mathrm{sum}}$, from their values obtained by exact diagonalization, $\langle\mathcal{O}\rangle_{\mathrm{exact}}$. 
This comparison shows the intrinsic error of the RBM parametrization and illustrates that it dominates the overall error for $N\geq 4$.}
  \label{fig:9}
\end{figure*}

We only consider moderate system sizes, $N\leq 10$, so that we can apply exact diagonalization, enabling an exact evaluation of any desired operator, and thus a benchmark of the dNN ansatz in arbitrary bases. We remark that, for these system sizes, sampling of configurations, usually used for evaluating gradients during training, can be omitted in favor of calculating the gradients exactly by summing over all states. This makes the obtained representation more accurate.

\subsection{Results}
\label{sec:results-1}
When training the RBM to represent a ground state of the TFIM, it turns out that it is sufficient to choose the weights purely real since the Hamiltonian is stoquastic, i.e., all its off-diagonal elements in the $z$-basis are real and non-positive \cite{Kivlichan2015,Bravyi2008}.
We train a real RBM with as many hidden as visible neurons, $M=N$, to represent the ground state of the TFIM at the critical point, $h=1$, and fix the weights. 
We use the dNN setup according to \Fig{3}, together with the phase reweighting scheme with block Gibbs sampling to perform measurements of operators in the $z$- and $x$-directions. Here we choose the ground state at the quantum critical point where the entanglement entropy is maximal and grows logarithmically with system size \cite{Igloi2008}.
This demonstrates that quantum effects can be represented in the classical network ansatz. It should be noted that two kinds of imperfections are involved now. First, the representation of the ground state as an RBM is not exact (representation error) and, second, there will be statistical errors due to finite sample sizes (sampling error). We analyze both aspects with the main focus being on the statistical sampling errors.

Figure \ref{fig:9} shows the results for the sampling error, where we vary the system size from $N=2$ to $N=10$. Panel (a) shows the absolute deviations between performing the phase reweighting scheme in the dNN representation and summing over all configurations explicitly, evaluating operators in the $x$-basis using local operators. We study magnetizations ($\sigma_1^z$ and $\sigma_1^x$) and nearest-neighbor correlations in the $z$- and $x$-directions ($\sigma_1^z\sigma_2^z$ and $\sigma_1^x\sigma_2^x$). Restricting to the magnetizations of the first spin and the correlations between the first two spins suffices due to translation invariance, which is explicitly implemented in the structure of the RBM weights \cite{Carleo2017, Sohn2012}. We run the sampling ten times for each system size and average the outcomes, with shaded regions denoting the statistical fluctuations.

We find good convergence, especially in the $z$-basis, where the weights are purely real and the complex phases vanish. The absolute deviations go down proportional to $1/\sqrt{\#\ \mathrm{Samples}}$, as it is expected due to statistical arguments \cite{Caflisch1998}. The error is approximately independent of system size.
Considering measurements in the $x$-basis, we find larger deviations, which is reasonable as now the weights in the network also take imaginary values and phases need to be considered, which can cancel each other in the normalization factor $Z(\mathcal{W})$. If these phases fluctuate heavily, a sign problem can appear. This leads to divergences for too small sample sizes and requires an exponentially growing amount of samples to find convergence for increasing system sizes. An increase in the absolute deviation with growing system size is indeed observed [see lower panels of Fig.~\ref{fig:9}(a)]. However, we still find stable convergence to the exact solution as expected from statistical reasons up to $N=10$.

Figure \ref{fig:9}(b) shows the absolute deviations of the expectation values of magnetizations and correlations in the $x$- and $z$-directions using $10^8$ samples, from results when summing over all states explicitly as a function of system size. The underlying data is the same as in panel (a). Here we see more clearly that the absolute deviations do not depend on the system size for measurements in the $z$-basis, but they scale exponentially with the system size for measurements in the $x$-basis (mind the log-scale). In this case the ansatz performs inefficiently, as exponentially many samples are necessary when going to larger system sizes. However, in the present example the sample size is still much smaller than the number of possible states in the network, which here is $2^{5N}$ as we choose $M=N$. This enables simulations of slightly larger systems than with exact diagonalization using comparable resources.

Figure \ref{fig:9}(c) shows the representation error, i.\,e.\ the absolute deviations of the observables obtained using the RBM parametrization and summing over all states explicitly from the expected value calculated via exact diagonalization. We find that the deviations grow abruptly larger with increasing system size for $N\geq4$, which is probably due to the limited representational power of the network. While the state can be parametrized with good accuracy for $N<4$, it takes a form for $N\geq 4$ which cannot be represented that accurately with the RBM ansatz. The deviations saturate around $10^{-2}$ for large system sizes, which is still small. This shows that the weights trained in the RBM parametrization represent the exact ground state with good accuracy for the cases considered here. However, the deviations are mostly larger than the ones in panel (a), showing that the overall error is dominated by the representation error and larger sample sizes cannot improve the accuracy any further.

In summary, we find that the ground state of the TFIM can be represented well with the RBM parametrization. It can be sampled using the phase reweighting scheme in the dNN ansatz to perform measurements in the $x$-basis, but due to  exponentially scaling sample sizes it is limited to small system sizes.

\section{Bell and GHZ States}\label{sec:Bell}
We now apply our dNN approach with phase reweighting to a paradigmatic example of an entangled state, the Bell state \cite{Bell1964,Bell1966,Bell2004}, and its generalization to larger spin systems, the Greenberger-Horne-Zeilinger (GHZ) states \cite{Greenberger1989}. With these examples we intend to assess whether the network can capture genuine quantum features such as entanglement and non-locality manifest in the violation of Bell's inequalities.

\subsection{Model and dNN Representation}
We consider a Bell state of two spin-1/2 particles, also called a Bell pair (BP),
\begin{align}
  \label{eq:18}
  \ket{\Psi_{\mathrm{BP}}}=&\ \frac{1}{\sqrt{2}}\left(\ket{\uparrow\downarrow}+\ket{\downarrow\uparrow}\right).
\end{align}
This state has non-classical correlations in the sense that it violates Bell's inequality. For classical systems Bell's inequality cannot be violated under the assumptions of local realism \cite{Bell1964,Bell1966,Bell2004}. Specifically, we consider the well-known CHSH-inequality (named after Clauser, Horne, Shimony and Holt) \cite{Clauser1969,Clauser1974}.
A CHSH-inequality which is maximally violated by the considered state is given by
\begin{align}
  \label{eq:19}
  \begin{split}
    \left|\mathcal{B}\right|=&\ \sqrt{2}\left|\left<\sigma_1^x\otimes\sigma_2^x\right>-\left<\sigma_1^z\otimes\sigma_2^z\right>\right|\\
    \leq&\ 2.
  \end{split}
\end{align}
If the measured correlations exceed this bound, they cannot arise in a classical way. However, with the wave function of the Bell state we find
\begin{align}
  \label{eq:20}
  \mathcal{B}=&\ 2\sqrt{2}>2,
\end{align}
showing that the CHSH-inequality is violated. It has been shown that this is the maximum reachable value for a quantum state, so that the inequality is maximally violated \cite{Cirelson1980}.

\begin{table*}
\begin{tabular}{l|cccccc}
  \hspace*{1cm} &\ \ \ \ &Bell State&\ \ \ \ &Bell State&\ \ \ \ &GHZ State\\
  &&(complex weights)&&(imaginary weights)&&\\
  \hline\\
  $d_j$&&$i\frac{\pi}{2}$ for $j=1$,& &$0$& &$0$\\
  &&$0$ otherwise&&&&\\ \\
  $b_k$&&$i\frac{\pi}{2}$& &$0$& &$i\frac{\pi}{2}$\\ \\
  &&&&&&$\frac{i}{2\left(N-1\right)}\mathrm{arcsin}\left[\frac{1}{2^{N-1/2}}\right]$ if $j\neq N$, $k=1$;\\
  $W_{j,k}$&&$\frac{\left(-1\right)^{j}}{2}\mathrm{arsinh}\left(\frac{1}{\sqrt{8}}\right)+i\frac{\pi}{2}$&&$i\left[\frac{\left(-1\right)^{j}}{2}\mathrm{arccos}\left(\frac{1}{\sqrt{8}}\right)-\frac{\pi}{4}\right]$&& $\frac{i}{2}\mathrm{arcsin}\left[\frac{1}{2^{N-1/2}}\right]$ if $j=N$, $k=1$;\\
  &&&&&& $i\frac{\pi}{4}\left(\delta_{j,k-1}+\delta_{j,k}\right)$ if $k\neq 1.$
\end{tabular}
\caption{Analytically calculated weights and biases entering the RBM representation of the Bell and GHZ states, where two possible solutions are given for the Bell state, one with complex values and one with purely imaginary entries. 
To avoid confusion with the imaginary unit we have used the indices $j,k$ instead of the usual convention $i,j$ to label the visible and hidden units, respectively.}
\label{tab:1}
\end{table*}

To represent the Bell state in the RBM, we need $N=2$ visible neurons and we show in \App{bell-pair-state} that it is sufficient to choose $M=1$ hidden neuron. Expressions for the weights in the RBM can be derived analytically by demanding
\begin{align}
  \label{eq:21}
  \begin{split}
    c_{\vec{v}^z=\left[\pm1,\pm1\right]}\left(\mathcal{W}\right)\stackrel{!}{=}&\ 0,\\
    c_{\vec{v}^z=\left[\pm1,\mp1\right]}\left(\mathcal{W}\right)\stackrel{!}{=}&\ \frac{1}{\sqrt{2}},
  \end{split}
\end{align}
see \App{bell-pair-state} for a full derivation and \Tab{1} for possible analytical expressions of the weights. Solving \Eq{21} yields infinitely many possible choices to represent a Bell state.
Thus we can directly determine the weights in the RBM and do not need to train them. We can rather perform sampling via the phase reweighting scheme to measure the correlations in the $x$- and $z$-directions and see if Bell correlations can be captured with this ansatz.

The generalization of the Bell state to larger system sizes yields the GHZ state with state vector \cite{Greenberger1989}
\begin{align}
  \label{eq:25}
  \ket{\Psi_{\mathrm{GHZ}}}=&\ \frac{1}{\sqrt{2}}\left(\ket{\uparrow\uparrow\dots\uparrow}+\ket{\downarrow\downarrow\dots\downarrow}\right).
\end{align}
This is a genuinely $N$-partite entangled state which we can represent in the dNN approach. As the GHZ state is the generalization of the Bell state to larger system sizes in the sense of being a superposition between two macroscopically different states, we use it to check the scalability of the dNN ansatz \cite{Greenberger1989,Dur2000,Gisin1998}.
The weights to represent this GHZ state with an RBM can be calculated analytically by solving
\begin{align}
  \label{eq:26}
  \begin{split}
    c_{\vec{v}^z=\left[\pm1,\dots,\pm1\right]}\left(\mathcal{W}\right)\stackrel{!}{=}\ &\frac{1}{\sqrt{2}},\\
    c_{\vec{v}^z\neq\left[\pm1,\dots,\pm1\right]}\left(\mathcal{W}\right)\stackrel{!}{=}&\ 0,
  \end{split}
\end{align}
which is solvable for $M=N-1$ hidden neurons, see \App{GHZ-state} for a full derivation. We derive a formula for the weights and biases in a general way as a function of the system size $N$. The final expressions for general system sizes are stated in \Tab{1}. 

\subsection{Results}
\label{sec:results}
We implement an RBM representing a Bell state, consisting of one hidden and two visible neurons and perform the phase reweighting scheme to sample from the underlying Boltzmann distribution using block Gibbs sampling \cite{Hinton2012}. We consider two possible choices for the network parameters, one with purely imaginary and one with complex weights. The expressions for the weights are given explicitly in the left two columns of \Tab{1}.

\begin{figure}[t]
  \centering
  \includegraphics[width=0.85\linewidth]{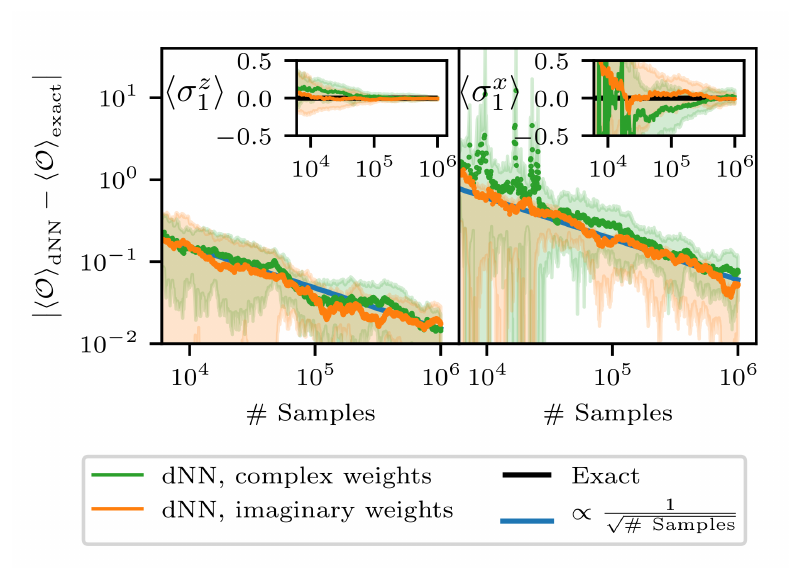}
  \caption{RBM representation of the Bell state \eq{18}:   
  Sampling errors for the phase reweighting scheme. 
  The magnetization of one spin of a Bell pair in the $x$- and $z$-directions is evaluated.
  The graphs show the absolute difference between the exact expectation values and those sampled from the dNN by means of the phase reweighting scheme, as a function of sample size. 
  Two choices of weights are compared, taking either purely imaginary (orange data) or complex values (green data). 
  The blue line indicates the deviation expected on statistical grounds.  
  Insets show the sampled values together with the respective exact solutions. 
  Ten simulation runs are averaged. 
  The shaded regions indicate the statistical fluctuations. 
  For the magnetization in the $x$-basis, the deviations decay only slowly as function of the sample size while fluctuations resulting from the sum over the complex phases are found for small samples.}
  \label{fig:5}
\end{figure}

As the system consists only of two sites, we can solve it exactly and compare the simulation outcome with the exact solution. The expected magnetizations and correlations in the $x$- and $z$-directions are given by
\begin{align}
  \label{eq:27}
  \begin{split}
    \left<\sigma_i^x\right>=\left<\sigma_i^z\right>=&\ 0,\ \forall\ i\in\left\{1,2\right\},\\
    \left<\sigma_1^x\sigma_2^x\right>=-\left<\sigma_1^z\sigma_2^z\right>=&\ 1.
  \end{split}
\end{align}
The simulation results of the magnetization in the $x$- and $z$-basis are shown in \Fig{5}, where for symmetry reasons we only plot the result for the first spin. Shown are results for both choices of the network parameters. The insets depict the corresponding observables together with the exact solutions as functions of sample size, and the main plots show the absolute deviations from the exact result. To compare the convergence to the expectation on grounds of statistical arguments \cite{Caflisch1998} we also display the expected sampling error $\sigma[\langle\mathcal{O}\rangle]/\sqrt{\#\ \mathrm{Samples}}$ where $\sigma^2[\langle\mathcal{O}\rangle]$ is the variance. The sampling error can be obtained via error propagation \cite{Caflisch1998},
\begin{align}
  \label{eq:36}
  \left|\left<\mathcal{O}\right>_{\mathrm{dNN}}\right.&\left.-\left<\mathcal{O}\right>_{\mathrm{exact}}\right|=\frac{\sigma\left[\left<\mathcal{O}\right>_{\mathrm{dNN}}\right]}{\sqrt{\#\ \mathrm{Samples}}},\\
  \label{eq:40}
  \begin{split}
    \sigma&\left[\left<\mathcal{O}\right>_{\mathrm{dNN}}\right]=\ \left|\frac{\sigma\left[\mathrm{Re}\left<\Psi\left|\mathcal{O}\right|\Psi\right>_{\mathrm{dNN}}\right]}{\mathrm{Re}\left<\Psi\left|\Psi\right.\right>_{\mathrm{dNN}}}\right|\\
    &\ +\left|\frac{\mathrm{Re}\left<\Psi\left|\mathcal{O}\right|\Psi\right>_{\mathrm{dNN}}}{\mathrm{Re}\left<\Psi\left|\Psi\right.\right>_{\mathrm{dNN}}^2}\sigma\left[\mathrm{Re}\left<\Psi\left|\Psi\right.\right>_{\mathrm{dNN}}\right]\right|,
  \end{split}
\end{align}
where the index ``exact'' refers to the exact quantum mechanical expectation value. The index ``dNN'' denotes the average over samples drawn from the dNN.

For such small system sizes as we consider them here, these expressions can be evaluated explicitly by summing over all possible network states. Due to the exact sum over all states the imaginary parts of the expectation values vanish with good accuracy and we hence neglect them. We expect the simulations to follow this decay for a sufficiently large number of samples, when the effects due to fluctuating phases are suppressed. In \Fig{5}, the blue line in the main plot denotes the expected convergence according to the explicitly evaluated variance for the case of purely imaginary weights. The expected convergence behavior for the case of complex weights shows a similar decay.
We find that the absolute deviation follows the expected sampling error accurately and hence converges to the exact solution, where no clear difference can be observed between the two cases of complex and purely imaginary weights.

\begin{figure}[t]
  \centering
  \includegraphics[width=\linewidth]{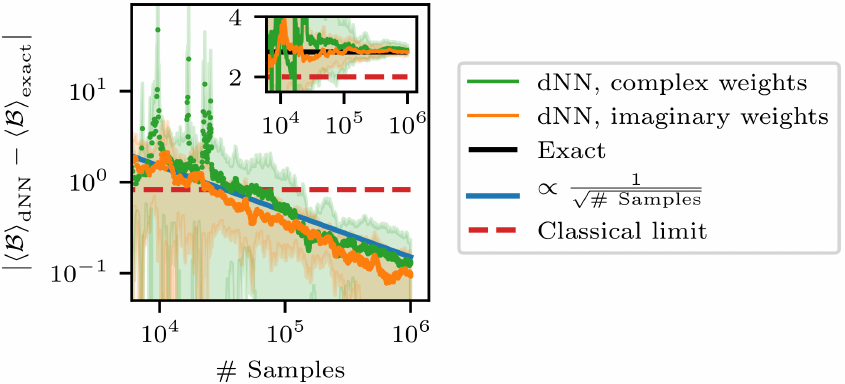}
  \caption{RBM representation of the Bell state:   
  observable $\mathcal{B}$ entering the CHSH Bell inequality \eq{19}.
  Shown is the same deviation as in \Fig{5}, as a function of samples drawn from the dNN representation of a Bell state with phase reweighting ansatz. 
  The weights can be chosen either complex or purely imaginary, indicated as green and orange data, respectively. 
  The main plot shows the absolute deviation of the simulation from the exact solution together with the expected convergence behavior, while the inset shows the direct evaluation of the observables in comparison to the exact solution. 
  Shaded regions indicate statistical fluctuations resulting from averaging over ten  runs.}
  \label{fig:6}
\end{figure}

To benchmark the phase reweighting scheme on the evaluation of correlations, we directly consider the CHSH-observable, \Eq{19}. The result is shown in \Fig{6}, which has the same structure as the plots in \Fig{5}.

In \Fig{6} we add a dashed line at the classical limit of $\mathcal{B}=2$ in the inset and at the deviation of the limit from the exact solution, $2\sqrt{2}-2$, in the main plot. The CHSH-inequality is hence violated if the curve in the inset is above and the curve in the main plot is below the dashed line.
We find clear convergence in agreement with the explicitly calculated behavior expected from statistical arguments. We also find a violation of the CHSH-inequality after rather short sampling times. However, we already consider $10^6$ samples here, which is large compared to the number of network configurations of the dNN which is $2^{N+2M}=2^{4}$ or $2^{N+2N+2M}=2^{8}$ for measurements in the $z$- or $x$-basis, respectively.

\begin{figure}
  \centering
  \includegraphics[width=\linewidth]{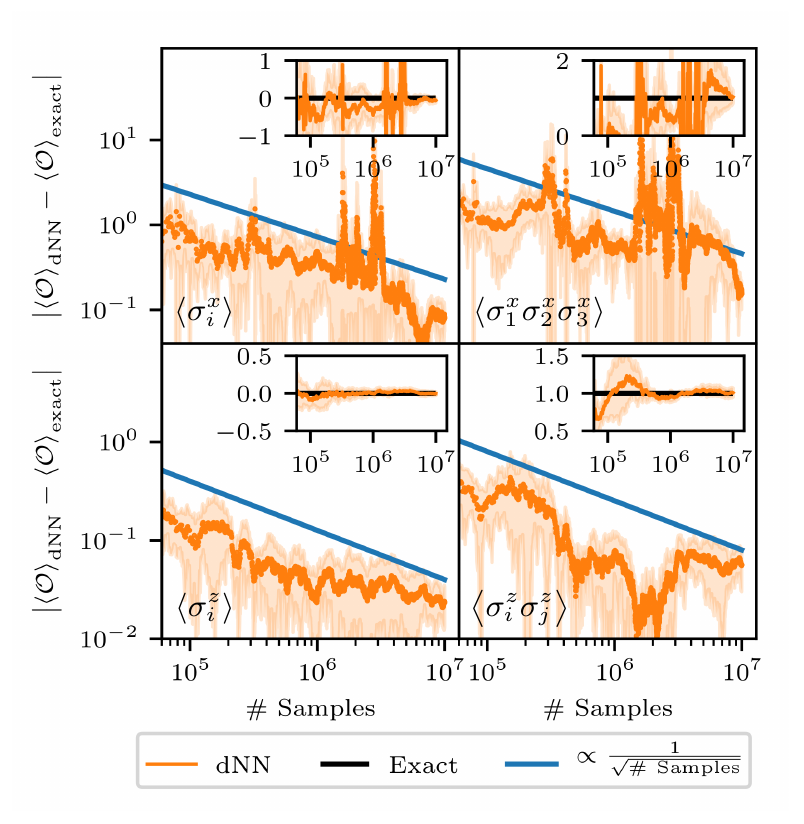}
  \caption{RBM representation of the GHZ state with $N=3$ spins:
  Magnetizations and correlations in the $x$- and $z$-bases resulting from sampling by means of the phase reweighting scheme as functions of sample size. 
  Insets show direct evaluations of the operators together with the exact outcome, while main plots show the absolute deviations from the exact solution compared to the explicitly evaluated expected convergence behavior. 
  Five simulation runs are averaged over, with statistical fluctuations indicated by the shaded regions. 
  Additionally, we average over the spin sites $i$ and $j$ due to translation invariance.}
  \label{fig:7}
\end{figure}

To analyze how the sample size necessary to find convergence scales with the system size we now consider GHZ states with $N>2$. We first consider the GHZ state for $N=3$ sites. The expected magnetizations and correlations are
\begin{align}
  \label{eq:28}
  \begin{split}
    \left<\sigma_i^x\right>=\left<\sigma_i^z\right>=&\ 0,\\
    \left<\sigma_1^x\sigma_2^x\sigma_3^x\right>=\left<\sigma_i^z\sigma_j^z\right>=&\ 1,\\
    \forall\ i,j\in\left\{1,2,3\right\}&.
  \end{split}
\end{align}
We set up an RBM with $M=2$ hidden neurons and the analytically derived,  purely imaginary weights, see the right column of \Tab{1} for explicit expressions, and perform the phase reweighting scheme in combination with block Gibbs sampling to measure magnetizations and correlations in the $x$- and $z$-directions. The results are shown in \Fig{7}, where each panel has the same structure as the plots for the Bell state.
We find convergence to the exact solution with the expected dependence on the sample size. Observables in the $z$-basis show faster convergence due to the smaller network size compared to off-diagonal observables which require an additional hidden layer. The expected convergence behavior can again be evaluated explicitly, as for the Bell state, since the network size is still small. For measurements in the $x$-basis we find huge fluctuations for small sample sizes, which basically result from the division by the sum over the phases, see \Eqs{10}{11}. This can lead to divergences due to cancellations of the phases. We would expect these fluctuations to vanish for larger sample sizes, however we already consider $10^7$ samples here, exceeding the number of network configurations of $2^7$ and $2^{13}$ for representing the state in the $z$- and $x$-basis, respectively. Besides the fluctuations, the functions decay as expected from statistical arguments. We even find the decay in the simulations slightly below the expected behavior, but a convergence to the blue curve is visible for large sample sizes.

\begin{figure}
  \centering
  \includegraphics[width=\linewidth]{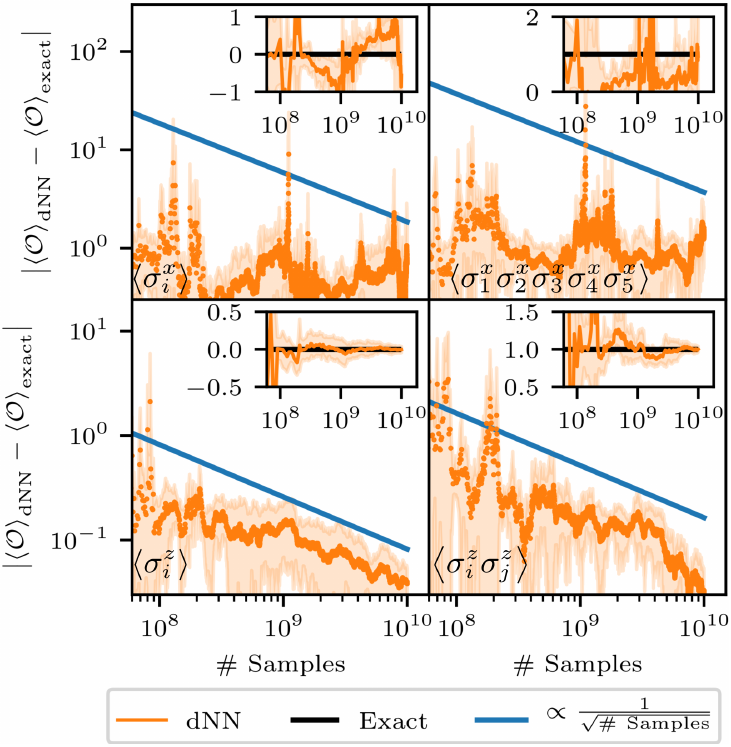}
  \caption{RBM representation of the GHZ state with $N=5$ spins:
  Magnetizations and correlations resulting in the phase reweighting scheme as functions of sample size. 
  Main plots show absolute deviations of the simulations from the exact solution together with the explicitly evaluated expected convergence behavior, while insets show the direct evaluations of the operators together with the exact solution. 
  Shaded regions denote statistical fluctuations resulting from averaging over five simulation runs as well as over the spin sites $i$ and $j$ due to translation invariance.}
  \label{fig:8}
\end{figure}
We now increase the system size further and consider a GHZ state with $N=5$ spins. We can represent it using an RBM with $M=4$ hidden neurons, see \App{GHZ-state}. The exact magnetizations and correlations are 
\begin{align}
  \label{eq:29}
  \begin{split}
    \left<\sigma_i^x\right>=\left<\sigma_i^z\right>=\ &0,\\
    \left<\sigma_1^x\sigma_2^x\sigma_3^x\sigma_4^x\sigma_5^x\right>=\left<\sigma_i^z\sigma_j^z\right>=\ &1,\\
    \forall\ i,j\in\left\{1,2,3,4,5\right\}&.
  \end{split}
\end{align}
The simulation results for these observables are shown in \Fig{8}, where we find convergence for measurements in the $z$-basis according to the expected statistical behavior. This can still be evaluated explicitly for the small network size. However, considering the outcome of measurements in the $x$-basis we find large fluctuations which do not decrease for the sample sizes considered. This indicates that we are still undersampling and convergence to the exact solution is not yet visible. According to the explicitly calculated error decay, we would expect convergence appearing for larger sample sizes. As we already consider $10^{10}$ samples, we could not increase the sample size further with the given computational setup. The networks here have $2^{13}$ and $2^{23}$ possible configurations to represent the system in the $z$- and $x$-basis, respectively.

In summary we find that the necessary sample size required to find convergence scales exponentially with the system or network size. This is due to the appearance of a sign problem, as it is known to be present for phase reweighting schemes in quantum Monte Carlo approaches \cite{Nakamura1992, Troyer2005, Anagnostopoulos2002, Loh1990}. Intuitively the sign problem  results from the sum over the phase factors, which can be distributed broadly on the unit circle in the complex plane. Thus, exponentially many samples are necessary to sum those phases up in the right way and get a stable result.

\section{Conclusion}\label{sec:conclusion}
When parametrizing wave functions of quantum spin systems with restricted Boltzmann machines, the weights and biases need to be chosen complex. Thus, the wave function can no longer be viewed as the marginal of a Boltzmann distribution but is rather a sum over complex terms \cite{Carleo2017, Carleo2019, Torlai2016, Saito2017}. This prohibits the straight forward implementation of the sampling from quantum states using classical neuromorphic structures.
We recover a way to sample the visible and hidden neurons from Boltzmann distributions defined by the real parts of the weights and biases via a phase reweighting scheme, where standard Gibbs sampling can be applied \cite{Hinton2012}. This enables an extension to deep neural networks with multiple layers, which we use to derive a representation of spin states in arbitrary bases.

When benchmarking this ansatz on highly entangled spin systems, we find an exponential scaling of the sample size necessary for convergence to the exact solution with increasing system or network size. This can be understood as a sign problem, meaning that the variances of the sampled quantities increase exponentially with system size. The method is hence rendered inefficient for cases where the wave-function coefficients cannot be chosen real and positive \cite{Nakamura1992,Troyer2005,Anagnostopoulos2002,Loh1990}.
However, when representing the ground state of the TFIM at the quantum critical point, we find that the phase reweighting scheme in the dNN ansatz yields accurate results for operators in the $z$-basis without dependence on the system size. For measurements in the $x$-basis we find an inefficient exponential scaling of the sample size with the system size.
In summary, our ansatz yields a generalization of the RBM parametrization of wave functions to deep networks and enables an implementation of the sampling on neuromorphic hardware, which can efficiently sample from Boltzmann distributions and could provide a  speedup and thus shift the limitations to larger system sizes, while it is not expected to in general overcome the curse of dimensionality of the quantum many-body problem \cite{Petrovici2016, Petrovici2016a, Kungl2019}.

\section*{Acknowledgments}
The authors thank A. Baumbach and L. Kades for discussions and collaborations on the topics described here. The authors acknowledge inspiring discussions with the late K. Meier. This work was supported by Deutsche Forschungsgemeinschaft (DFG) under the SFB 1225 (ISOQUANT) and under Germany's Excellence Strategy EXC-2181/1-390900948 (the Heidelberg STRUCTURES Excellence Cluster), by Heidelberg University and by the state of Baden-W\"urttemberg through bwHPC.

\makeatletter
\appendix
\section{Representation of the Bell State}
\label{app:bell-pair-state}
To parametrize the Bell state with an RBM, we need $N=2$ visible neurons for the two spins and it is sufficient to choose $M=1$ hidden neuron. We can analytically derive expressions for the weights to represent the corresponding basis expansion coefficients. The state vector reads \cite{Bell1964,Bell1966,Bell2004}
\begin{align}
  \label{eq:91}
  \begin{split}
    \ket{\Psi_{\mathrm{BP}}}=&\ \frac{1}{\sqrt{2}}\left(\ket{\uparrow\downarrow}+\ket{\downarrow\uparrow}\right)\\
    =&\sum_{\left\{\vec{v}^z\right\}}c_{\vec{v}^z}\left(\mathcal{W}\right)\ket{\vec{v}^z}.
  \end{split}
\end{align}
Enumerating the two states as $|\downarrow\rangle=|-1\rangle$ and $|\uparrow\rangle=|1\rangle$, we get
\begin{align}
  \label{eq:92}
  \begin{split}
    c_{\vec{v}^z=\left(\pm1,\pm1\right)}\left(\mathcal{W}\right)=&\ \mathrm{exp}\left[\pm\left(d_1+d_2\right)\right]\\
    &\times 2\mathrm{cosh}\left[\pm\left(W_{1,1}+W_{2,1}\right)+b_1\right]\\
  \stackrel{!}{=}&\ 0,
  \end{split}\\
  \label{eq:93}
  \begin{split}
    c_{\vec{v}^z=\left(\pm1,\mp1\right)}\left(\mathcal{W}\right)=&\ \mathrm{exp}\left[\pm\left(d_1-d_2\right)\right]\\
    &\times 2\mathrm{cosh}\left[\pm\left(W_{1,1}-W_{2,1}\right)+b_1\right]\\
  \stackrel{!}{=}&\ \frac{1}{\sqrt{2}}.
  \end{split}
\end{align}
Here we use the RBM parametrization of the coefficients, see \Eqs{2}{3} in the main text,
\begin{align}
  \label{eq:94}
  \begin{split}
  c_{\vec{v}^z}\left(\mathcal{W}\right)=&\ \mathrm{exp}\left[\sum_{i=1}^{N}d_iv_i^z\right]\\
  &\times\prod_{j=1}^M 2\mathrm{cosh}\left[\sum_{i=1}^N v_i^zW_{i,j}+b_j\right].
  \end{split}
\end{align}
From \Eq{92} we choose the ansatz
\begin{align}
  \label{eq:95}
  \mathrm{cosh}\left[\pm\left(W_{1,1}+W_{2,1}\right)+b_1\right]\stackrel{!}{=}&\ 0\\
  \Rightarrow b_1\pm\left(W_{1,1}+W_{2,1}\right)\stackrel{!}{=}&\ i\pi\left(n_{\pm}+\frac{1}{2}\right),\ \ \ n_{\pm}\in\mathbb{Z},\\
  \label{eq:96}
  \begin{split}
    \Rightarrow b_1=&\ i\pi\left(\frac{n_++n_-}{2}+\frac{1}{2}\right),\\
    W_{1,1}+W_{2,1}=&\ i\pi\frac{n_+-n_-}{2}.
  \end{split}
\end{align}
Inserting this into \Eq{93}, we get two expressions depending on whether $n^{(\pm)}=n_+\pm n_-$ is even or odd,
\begin{align}
  \label{eq:97}
  \begin{split}
    \mathrm{cosh}&\left[b_1\pm\left(W_{1,1}-W_{2,1}\right)\right]\\
    =&\begin{cases}\pm i\left(-1\right)^{n^{\left(+\right)}/2}\mathrm{sinh}\left[W_{1,1}-W_{2,1}\right],\ \mathrm{if}\ n^{\left(\pm\right)}\ \mathrm{even},\\ \left(-1\right)^{\left(n^{\left(+\right)}+1\right)/2}\mathrm{cosh}\left[W_{1,1}-W_{2,1}\right],\ \mathrm{if}\ n^{\left(\pm\right)}\ \mathrm{odd}.\end{cases}
  \end{split}
\end{align}
Considering the even case, \Eqs{92}{93} turn into
\begin{align}
  \label{eq:98}
  \begin{split}
    i\left(-1\right)^{n^{\left(+\right)}/2}\mathrm{sinh}\left[d_1-d_2\right]\mathrm{sinh}\left[W_{1,1}-W_{2,1}\right]\stackrel{!}{=}&\ 0,\\
    i\left(-1\right)^{n^{\left(+\right)/2}}\mathrm{cosh}\left[d_1-d_2\right]\mathrm{sinh}\left[W_{1,1}-W_{2,1}\right]\stackrel{!}{=}&\ \frac{1}{\sqrt{8}}.
  \end{split}
\end{align}
From this, it follows that
\begin{align}
  \label{eq:100}
  \begin{split}
    d_1-d_2=&\ i\pi\left(m_{\mathrm{e}}+\frac{1}{2}\right),\ \ m_{\mathrm{e}}\in\mathbb{Z},\\
    W_{1,1}-W_{2,1}=&\left(-1\right)^{n^{\left(+\right)}/2+m_{\mathrm{e}}+1}\mathrm{arsinh}\left[\frac{1}{\sqrt{8}}\right].
  \end{split}
\end{align}
Together with \Eq{96} we get solutions for even $n^{(\pm)}$,
\begin{align}
  \label{eq:102}
  \begin{split}
    W_{1,1}=&\left(-1\right)^{n^{\left(+\right)}/2+m_{\mathrm{e}}+1}\frac{1}{2}\mathrm{arsinh}\left[\frac{1}{\sqrt{8}}\right]+i\frac{\pi}{4}n^{\left(-\right)},\\
    W_{2,1}=&\left(-1\right)^{n^{\left(+\right)}/2+m_{\mathrm{e}}}\frac{1}{2}\mathrm{arsinh}\left[\frac{1}{\sqrt{8}}\right]+i\frac{\pi}{4}n^{\left(-\right)},\\
    d_1-d_2=&\ i\pi\left(m_{\mathrm{e}}+\frac{1}{2}\right),\\
    b_1=&\ i\pi\left(\frac{n^{\left(+\right)}}{2}+\frac{1}{2}\right).
  \end{split}
\end{align}
For the case of odd $n^{(\pm)}$, \Eqs{92}{93} yield
\begin{align}
  \label{eq:103}
  \begin{split}
    \left(-1\right)^{\left(n^{\left(+\right)}+1\right)/2}\mathrm{cosh}\left[d_1-d_2\right]\mathrm{cosh}\left[W_{1,1}-W_{2,1}\right]\stackrel{!}{=}&\ \frac{1}{\sqrt{8}},\\
    \left(-1\right)^{\left(n^{\left(+\right)}+1\right)/2}\mathrm{sinh}\left[d_1-d_2\right]\mathrm{cosh}\left[W_{1,1}-W_{2,1}\right]\stackrel{!}{=}&\ 0.
  \end{split}
\end{align}
This leads to
\begin{align}
  \label{eq:105}
  \begin{split}
    d_1-d_2=&\ i\pi m_{\mathrm{o}},\ \ m_{\mathrm{o}}\in\mathbb{Z},\\
    W_{1,1}-W_{2,1}=&\ i\left[\pi\left(\frac{n^{\left(+\right)}+1}{2}+m_{\mathrm{o}}\right)\right.\\
    &\left.+\left(-1\right)^{\left(n^{\left(+\right)}+1\right)/2+m_{\mathrm{o}}}\mathrm{arccos}\left(\frac{1}{\sqrt{8}}\right)\right].
  \end{split}
\end{align}
So we get for odd $n^{(\pm)}$ the solutions
\begin{align}
  \label{eq:107}
  \begin{split}
    W_{1,1}=&\ i\left[\left(-1\right)^{\left(n^{\left(+\right)}+1\right)/2+m_{\mathrm{o}}}\frac{1}{2}\mathrm{arccos}\left(\frac{1}{\sqrt{8}}\right)\right.\\
    &\left.+\frac{\pi}{4}\left(n^{\left(+\right)}+n^{\left(-\right)}+1+2m_{\mathrm{o}}\right)\right],\\
    W_{2,1}=&\ i\left[\left(-1\right)^{\left(n^{\left(+\right)}+1\right)/2+m_{\mathrm{o}}+1}\frac{1}{2}\mathrm{arccos}\left(\frac{1}{\sqrt{8}}\right)\right.\\
    &\left.+\frac{\pi}{4}\left(n^{\left(-\right)}-n^{\left(+\right)}-1-2m_{\mathrm{o}}\right)\right],\\
    d_1-d_2=&\ i\pi m_{\mathrm{o}},\\
    b_1=&\ i\frac{\pi}{2}\left(n^{\left(+\right)}+1\right).
  \end{split}
\end{align}
The solutions are highly degenerate, as three integers can be chosen arbitrarily, so we find infinitely many possibilities to choose the weights. In the main text we focus on two specific choices, one with complex and one with purely imaginary weights. The expressions for complex weights are given by choosing $n_+=1$, $n_-=-1$, so that we consider the case of even $n^{(\pm)}$. Additionally we choose $m_e=0$, yielding
\begin{align}
  \label{eq:30}
  \begin{split}
    d_1-d_2=&\ i\frac{\pi}{2}\Rightarrow d_1=i\frac{\pi}{2},\ d_2=0,\\
    b_1=&\ i\frac{\pi}{2},\\
    W_{1,1}=&-\frac{1}{2}\mathrm{arsinh}\left(\frac{1}{\sqrt{8}}\right)+i\frac{\pi}{2},\\
    W_{2,1}=&\ \frac{1}{2}\mathrm{arsinh}\left(\frac{1}{\sqrt{8}}\right)+i\frac{\pi}{2}.
  \end{split}
\end{align}
The second case we consider in the main text is the choice of purely imaginary weights, which we reach by setting $n_+=-1$, $n_-=0$, so that $n^{(\pm)}$ is odd. By also setting $m_o=0$, we get
\begin{align}
  \label{eq:31}
  \begin{split}
    d_1-d_2=&\ 0\Rightarrow d_1=d_2=0,\\
    b_1=&\ 0,\\
    W_{1,1}=&\ i\left(-\frac{1}{2}\mathrm{arccos}\left[\frac{1}{\sqrt{8}}\right]-\frac{\pi}{4}\right),\\
    W_{2,1}=&\ i\left(\frac{1}{2}\mathrm{arccos}\left[\frac{1}{\sqrt{8}}\right]-\frac{\pi}{4}\right).
  \end{split}
\end{align}
These are the explicit values for the weights and biases as stated in \Tab{1} in the main text.\\

\section{Representing the GHZ-State}
\label{app:GHZ-state}
For a general spin-1/2 system with $N$ sites, the Greenberger-Horne-Zeilinger (GHZ) state is described by \cite{Greenberger1989}
\begin{align}
  \label{eq:22}
  \ket{\Psi_{\mathrm{GHZ}}}=&\ \frac{1}{\sqrt{2}}\left(\ket{\uparrow\uparrow\dots\uparrow}+\ket{\downarrow\downarrow\dots\downarrow}\right).
\end{align}
The GHZ state is a strongly entangled quantum state consisting of at least three spin-1/2 particles \cite{Dur2000, Gisin1998}. To derive the weights representing such a GHZ state in the RBM parametrization, we first consider the case of $N=3$ sites. We add $M=2$ hidden neurons to the neural network and consider the parametrization of the basis state expansion coefficients as stated in \Eq{3} in the main text. This provides a set of four equations,
\begin{widetext}
\begin{align}
  \label{eq:24}
  \begin{split}
    c\left(\vec{v}^z=\left[\pm 1,\pm 1,\pm 1\right];\mathcal{W}\right)=&\ \mathrm{exp}\left[\pm\left(d_1+d_2+d_3\right)\right]4\mathrm{cosh}\left[b_1\pm\left(W_{1,1}+W_{2,1}+W_{3,1}\right)\right]\mathrm{cosh}\left[b_2\pm\left(W_{1,2}+W_{2,2}+W_{3,2}\right)\right]\\
    \stackrel{!}{=}&\ \frac{1}{\sqrt{2}},
  \end{split}\\
  \label{eq:51}
  \begin{split}
    c\left(\vec{v}^z=\left[\pm 1,\pm 1,\mp 1\right];\mathcal{W}\right)=&\ \mathrm{exp}\left[\pm\left(d_1+d_2-d_3\right)\right]4\mathrm{cosh}\left[b_1\pm\left(W_{1,1}+W_{2,1}-W_{3,1}\right)\right]\mathrm{cosh}\left[b_2\pm\left(W_{1,2}+W_{2,2}-W_{3,2}\right)\right]\\
    \stackrel{!}{=}&\ 0,
  \end{split}
          \end{align}
  \begin{align}
  \label{eq:52}
  \begin{split}
    c\left(\vec{v}^z=\left[\pm 1,\mp 1,\pm 1\right];\mathcal{W}\right)=&\ \mathrm{exp}\left[\pm\left(d_1-d_2+d_3\right)\right]4\mathrm{cosh}\left[b_1\pm\left(W_{1,1}-W_{2,1}+W_{3,1}\right)\right]\mathrm{cosh}\left[b_2\pm\left(W_{1,2}-W_{2,2}+W_{3,2}\right)\right]\\
    \stackrel{!}{=}&\ 0,
  \end{split}\\
  \label{eq:53}
  \begin{split}
    c\left(\vec{v}^z=\left[\mp 1,\pm 1,\pm 1\right];\mathcal{W}\right)=&\ \mathrm{exp}\left[\pm\left(d_2+d_3-d_1\right)\right]4\mathrm{cosh}\left[b_1\pm\left(W_{2,1}+W_{3,1}-W_{1,1}\right)\right]\mathrm{cosh}\left[b_2\pm\left(W_{2,2}+W_{3,2}-W_{1,2}\right)\right]\\
    \stackrel{!}{=}&\ 0.
  \end{split}          
\end{align}
\end{widetext}
There are many ways to solve this set of equations, but in the following we only look for one possible solution. Thus, from \Eq{51} we choose
\begin{align}
  \label{eq:77}
  \begin{split}
    \mathrm{cosh}\left[b_1\pm\left(W_{1,1}+W_{2,1}-W_{2,2}\right)\right]\stackrel{!}{=}&\ 0.
    \end{split}
    \end{align}
    From this it follows,
    \begin{align}
    \begin{split}
    &\ b_1=i\pi\left(\frac{n_1^++n_1^-}{2}+\frac{1}{2}\right),\\
    &\ W_{1,1}+W_{2,1}-W_{3,1}=i\pi\frac{n_1^+-n_1^-}{2}.
  \end{split}
\end{align}
Analogously, we can demand from \Eq{52}
\begin{align}
  \label{eq:78}
  \begin{split}
    \mathrm{cosh}&\left[b_2\pm\left(W_{1,2}-W_{2,2}+W_{3,2}\right)\right]\stackrel{!}{=}0\\
    \Rightarrow b_2=&\ i\pi\left(\frac{n_2^++n_2^-}{2}+\frac{1}{2}\right),\ \ n_2^{\pm}\in\mathbb{Z}\\
    W_{1,2}&-W_{2,2}+W_{3,2}=i\pi\frac{n_2^+-n_2^-}{2},
  \end{split}
\end{align}
and from \Eq{53}
\begin{align}
  \label{eq:135}
  \begin{split}
    \mathrm{cosh}&\left[b_2\pm\left(-W_{1,2}+W_{2,2}+W_{3,2}\right)\right]\stackrel{!}{=}0\\
    \Rightarrow -W_{1,2}&+W_{2,2}+W_{3,2}=i\pi\frac{n_2^+-n_2^-}{2}.
  \end{split}
\end{align}
As we only look for a single possible solution, we consider the simplest case with $n_i^{\pm}=0$ for $i=1,2$. This yields
\begin{align}
  \label{eq:138}
  \begin{split}
    b_1=b_2=&\ i\frac{\pi}{2},\\
    W_{1,1}+W_{2,1}-W_{3,1}=&\ 0,\\
    W_{3,2}=&\ 0,\ W_{1,2}=W_{2,2}.
  \end{split}
\end{align}
Plugging these results into \Eq{24} and considering only the cosh-terms gives
\begin{align}
  \label{eq:139}
  \begin{split}
    \mathrm{cosh}&\left[i\frac{\pi}{2}\pm\left(W_{1,1}+W_{2,1}+W_{3,1}\right)\right]\mathrm{cosh}\left[i\frac{\pi}{2}\pm 2W_{1,2}\right]\\
    =&\left\{\pm i\mathrm{sinh}\left[W_{1,1}+W_{2,1}+W_{3,1}\right]\right\}\left\{\pm i\mathrm{sinh}\left[2W_{1,2}\right]\right\}\\
    =&-\mathrm{sinh}\left[W_{1,1}+W_{2,1}+W_{3,1}\right]\mathrm{sinh}\left[2W_{1,2}\right]\\
    =&-i\mathrm{sinh}\left[W_{1,1}+W_{2,1}+W_{3,1}\right],
  \end{split}
\end{align}
where we again simplify the expression by choosing $W_{1,2}=i\pi/4$ in the last line. With this choice, \Eq{24} becomes
\begin{align}
  \label{eq:140}
  \begin{split}
    -i\mathrm{exp}&\left[\pm\left(d_1+d_2+d_3\right)\right]\mathrm{sinh}\left[W_{1,1}+W_{2,1}+W_{3,1}\right]\\
    =&-i\left(\mathrm{cosh}\left[d_1+d_2+d_3\right]\pm i\mathrm{sinh}\left[a_1+a_2+a_3\right]\right)\\
    &\times\mathrm{sinh}\left[W_{1,1}+W_{2,1}+W_{3,1}\right]\\
    \stackrel{!}{=}&\ \frac{1}{4\sqrt{2}}
  \end{split}\\
  \label{eq:141}
  \begin{split}
    \Rightarrow&\ \mathrm{sinh}\left[d_1+d_2+d_3\right]\stackrel{!}{=}\ 0\quad
    \Rightarrow d_1+d_2+d_3=\ 0\\
    \Rightarrow&\ -i\mathrm{sinh}\left[W_{1,1}+W_{2,1}+W_{3,1}\right]\stackrel{!}{=}\frac{1}{4\sqrt{2}}\\
    \Rightarrow&\ W_{1,1}+W_{2,1}+W_{3,1}=i\mathrm{arcsin}\left[\frac{1}{4\sqrt{2}}\right].
  \end{split}
\end{align}
Given these conditions, we can choose one possible solution with the weights
\begin{align}
  \label{eq:145}
  \begin{split}
    d_1=d_2=d_3=&\ 0,\\
    b_1=b_2=&\ i\frac{\pi}{2},\\
    W_{1,1}=W_{2,1}=&\ \frac{i}{4}\mathrm{arcsin}\left[\frac{1}{4\sqrt{2}}\right],\\
    W_{3,1}=&\ \frac{i}{2}\mathrm{arcsin}\left[\frac{1}{4\sqrt{2}}\right],\\
    W_{1,2}=W_{2,2}=&\ i\frac{\pi}{4},\ W_{3,2}=0.
  \end{split}
\end{align}
These are the weights we choose in the main text for $N=3$ sites.

From these results we can see that the choice of the weights to the first hidden neuron guarantees the normalized coefficients if all spins are equal and the zero coefficients if the third spin is flipped compared to the other two. The weights to the second hidden neuron guarantee the zero coefficients for the case that spins one and two have opposite sign, so that all cases are covered. This behavior can be generalized to an arbitrary number $N$ of spins and yields the weights to represent a GHZ state in a neural network with $M=N-1$ hidden neurons,
\begin{align}
  \label{eq:151}
  \begin{split}
    d_j=&\ 0,\qquad
    b_k=\ i\frac{\pi}{2},\\
    W_{j\neq N,1}=&\ \frac{i}{2\left(N-1\right)}\mathrm{arcsin}\left[\frac{1}{2^{N-1/2}}\right],\\
    W_{N,1}=&\ \frac{i}{2}\mathrm{arcsin}\left[\frac{1}{2^{N-1/2}}\right],\\
    W_{j,k\neq 1}=&\ i\frac{\pi}{4}\left(\delta_{j,k-1}+\delta_{j,k}\right),\\
    &\ \ \forall j\in\left\{1,\dots,N\right\},\ k\in\left\{1,\dots,M\right\}.
  \end{split}
\end{align}
To avoid confusion with the imaginary unit, we have replaced the indices $i$, $j$ of the visible and hidden neurons by the indices $j$, $k$, respectively. These are the expressions quoted in \Tab{1} in the main text and it has been checked that this choice of weights represents the GHZ state for small systems,  $N\leq 6$. The weights for the calculations with $N=5$ sites in the main text are chosen according to these equations.

\bibliographystyle{apsrev4-1}

\end{document}